\documentclass[12pt,reqno]{amsart}

\usepackage{amsmath}
\usepackage{amsfonts}
\usepackage{amsthm}
\usepackage{amstext}

\newcommand{\R}{{\mathbb R}}

\theoremstyle{plain}
\newtheorem{thm}{Theorem}[section]
\newtheorem{lemma}[thm]{Lemma}
\theoremstyle{definition}
\newtheorem*{acknowledgement}{Acknowledgement} 
\newtheorem{remark}[thm]{Remark}{\it}{\rm}

\newenvironment{pf*}[1]{\par\medskip\noindent\textit{#1}\,:}{\hspace*{\fill}\qed\medskip\par\noindent}   

\numberwithin{equation}{section}

\title[Regularity of the density at nuclei]{Non-Isotropic Cusp
       Conditions and Regularity of the Electron Density 
       of Molecules at the Nuclei} 

\thanks{\copyright\ 2006 by the
       authors. This article may be reproduced in its entirety for
       non-commercial purposes.}

\author[S. Fournais, M. and T. Hoffmann-Ostenhof, and T. \O. S\o rensen]
{S. Fournais \and M. Hoffmann-Ostenhof \and T. Hoffmann-Ostenhof \and
T. \O stergaard S\o rensen}

\address[S. Fournais]{CNRS and Laboratoire de Math\'{e}matiques,
           Universit\'{e} Paris-Sud - B\^{a}t 425,
           F-91405 Orsay Cedex, France.
           }
\email{soeren.fournais@math.u-psud.fr}

\address[T. \O stergaard S\o rensen]
{Institut des Hautes \'Etudes Scientifiques, Le Bois-Marie, 35, route
  de Chartres, F-91440 Bures-sur-Yvette, France} 

\address[T. \O stergaard S\o rensen, permanent address]
{Department of Mathematical Sciences,
           Aalborg University,
           Fredrik Bajers Vej 7G,
           DK-9220 Aalborg East, Denmark.}
\email{sorensen@math.aau.dk}

\address[M. Hoffmann-Ostenhof]
        {Fakult\"at f\"ur Mathematik,
         Universit\"at Wien,         
         Nordbergstra\ss e 15,
         A-1090 Vienna,
         Austria.} 
\email{maria.hoffmann-ostenhof@univie.ac.at}

\address[T. Hoffmann-Ostenhof]{Institut f\"ur Theoretische
Chemie, W\"ahringer\-strasse 17,
           Universit\"at Wien,
           A-1090 Vienna,
           Austria.}

\address[T. Hoffmann-Ostenhof, 2nd address]{
        The Erwin Schr\"{o}dinger International Institute for 
        Mathematical Physics,
              Boltzmanngasse 9,
              A-1090 Vienna, Austria.}
\email{thoffman@esi.ac.at}

\date{\today}

\begin{document}

\thispagestyle{empty}

\begin{abstract}
We investigate regularity properties  of molecular one-electron
densities $\rho$ near the nuclei. In particular we derive a
re\-pre\-sentation $$\rho(x)=e^{\mathcal F(x)}\mu(x)$$ with an
explicit function $\mathcal F$, only depending on the nuclear 
charges and the positions of the nuclei, such that $\mu\in
C^{1,1}(\mathbb R^3)$, i.e., $\mu$ has locally essentially bounded second
derivatives. An example constructed using Hydrogenic eigenfunctions
shows that this regularity result is sharp.  
For atomic eigenfunctions which are either even or odd with respect to 
inversion in the origin, we prove that \(\mu\) is even 
\(C^{2,\alpha}(\R^3)\) for all \(\alpha\in(0,1)\). 
Placing one nucleus at the origin we study  $\rho$ in polar
coordinates  $x=r\omega$ and investigate $\frac{\partial}{\partial
  r}\rho(r,\omega)$ and $\frac{\partial^2}{\partial r^2}\rho(r,\omega)$
for fixed $\omega$ as $r$ tends to zero.
We prove non-isotropic cusp conditions of first and second order,
which generalize Kato's classical result.
\end{abstract}

\maketitle

\section{Introduction and statement of the results}
We consider a non-relativistic  $N$-electron molecule with the nuclei
fixed in \(\R^3\). The Hamiltonian describing the system is given by 
\begin{equation}\label{Hbis}
  H=\sum_{j=1}^N\Big(-\Delta_j-\sum_{k=1}^K\frac{Z_k}{|x_j-R_k|}\Big)
  +\sum_{1\le i<j\le N}\frac{1}{|x_i-x_j|}.  
\end{equation}
Here the $R_k$, $k=1,\dots,  K,$ $R_i\neq R_j$ for $i\neq j$,
denote the positions of the (fixed) nuclei in  
$\mathbb R^3$ with charges $Z_1,\ldots,Z_K$, and the
$x_j=(x_{j,1},x_{j,2},x_{j,3})\in \mathbb R^3$, 
$j=1,\dots, N$, denote  the positions of the  
electrons. The  $\Delta_j, \:j=1,\dots, N$, are the associated
Laplacians so that $\Delta=\sum_{j=1}^N\Delta_j$  
is the $3N$-dimensional Laplacian. 
Let $\mathbf x=(x_1,x_2,\dots, x_N)\in \mathbb R^{3N}$ and
$\nabla=(\nabla_1,\dots, \nabla_N)$ denote the points in  
$\mathbb R^{3N}$ and the $3N$-dimensional gradient operator
respectively. We write  
$H=-\Delta+V$ where $V$ is the multiplicative potential 
\begin{equation}\label{Vpot}
  V({\bf x}) =-\sum_{j=1}^N
  \sum_{k=1}^K\frac{Z_k}{|x_j-R_k|}+\sum_{1\le i<j\le
    N}\frac{1}{|x_i-x_j|}. 
\end{equation}
Here we neglect the internuclear repulsion $W=\sum_{1\leq k<\ell\leq
  K}^K\frac{Z_kZ_\ell}{|R_k-R_\ell|}$ 
which is just an additive term in the fixed-nuclei approximation.

The operator $H$ is selfadjoint with operator domain $\mathcal
D(H)=W^{2,2}(\mathbb R^{3N})$ and form  
domain $\mathcal Q(H)=W^{1,2}(\mathbb R^{3N})$ \cite{kato2}. 

Let \(\psi\) be an eigenfunction of \(H\) corresponding to an
eigenvalue \(E\in\R\), that is, 
\begin{equation}\label{Hpsi}
  H\psi=E\psi.
\end{equation}
We shall here only consider eigenfunctions $\psi\in L^2(\mathbb
R^{3N})$\footnote{For scattering problems and  for solid state physics,
solutions to \eqref{Hpsi}  
which are not in $L^2$ are also important.}.

Note that physical molecular eigenfunctions have to satisfy the Pauli
principle. This is however irrelevant 
to our results and we impose no such condition.

The operator in \eqref{Hbis} (possibly with the addition of the 
internuclear repulsion \(W\)) can be considered as the standard model
for atoms and molecules in quantum mechanics. The analysis of  $H$ is
fundamental for the understanding of the properties of atoms,
molecules or, more generally, of matter\footnote{For some problems it
  is of course necessary to  
include nuclear motion, and in the presence of 
heavy nuclei relativistic effects have to be accounted
for.}.

It is well known that every eigenfunction $\psi$ of $H$ is
Lipschitz-conti\-nuous \cite{kato1} and real analytic  
away from the points in configuration space $\mathbb{R}^{3N}$ where
the potential $V$ defined in \eqref{Vpot} is singular (see
\cite[Section 7.5, pp. 177--180]{Hormander}). 
In this paper we investigate regularity properties of the electron
density $\rho$ associated to 
an eigenfunction $\psi$.  
The density $\rho$ is defined by 
\begin{equation}\label{rhohat}
  \rho(x)=\sum_{j=1}^N \rho_j(x)=
  \sum_{j=1}^N\int_{\mathbb
    R^{3N-3}}|\psi(x,\mathbf{\hat{x}}_j)|^2\,d\mathbf{\hat{x}}_j 
\end{equation}
where we use the notation 
\begin{align*}
  {\mathbf{\hat{x}}_j}:=(x_1,\dots,x_{j-1},x_{j+1},\dots, x_N),
\end{align*}
and
\begin{align*}
  d{\mathbf{\hat{x}}_j}:=dx_1\dots dx_{j-1}dx_{j+1}\dots dx_N,
\end{align*}
and, by abuse of notation, we identify
$(x_1,\dots,x_{j-1},x,x_{j+1},\dots,x_N)$ with
$(x,\mathbf{\hat{x}}_j)$.

We assume throughout when studying $\rho$ stemming from some
eigenfunction $\psi$ that  
\begin{equation}\label{eq:exp-dec}
  |\psi({\bf x})|\le C_0e^{-\gamma_0|\mathbf x|} \text{ for all }
  \mathbf x\in \mathbb R^{3N} 
\end{equation}
for some $C_0, \gamma_0>0$. 
By \cite[Theorem 1.2]{AHP} (see also \cite[Remark 1.7]{AHP}) this
implies the existence of constants 
\(C_{1}, \gamma_{1}>0\) such that
\begin{align}
  \label{eq:dec_grad_psi}
  \big|\nabla\psi(\mathbf x)\big|\leq C_{1}\,e^{-\gamma_{1}|\mathbf x|}
   \quad\text{ for almost all } \mathbf x\in\mathbb R^{3N}.
\end{align}
Since \(\psi\) is continuous,
\eqref{eq:exp-dec} is only an assumption on the behaviour at infinity. 
For references on the exponential decay of eigenfunctions, see e.g.\
Agmon~\cite{Agmon}, Froese and Herbst~\cite{froese-herbst}, and
Simon~\cite{simon1}.
 The proofs of our
results rely (if not indicated otherwise) on some kind of decay-rate
for \(\psi\); exponential decay is not essential, but assumed for
convenience.
Note that 
\eqref{eq:exp-dec} and \eqref{eq:dec_grad_psi} imply
that \(\rho\) is Lipschitz continuous in \(\R^3\) by Lebesgue's
theorem on dominated convergence.

In \cite{ArkMat} we showed that $\rho$ is real analytic away from
the nuclei ($\rho\in C^{\omega}(\mathbb R^3\setminus\{R_1,\dots,
R_K\})$); for earlier results see also \cite{AHP}, 
\cite{CMP1} and  \cite{COM}.  Note that the proof of the
analyticity does {\em not} require any decay of $\psi$ (apart from
$\psi\in W^{2,2}(\mathbb R^{3N})$). That $\rho$ itself is not analytic
in all of $\R^3$ is already clear for the ground state of the Hydrogen
atom ($N=K=1; R_1=0, Z_1=1$):
$\psi(x)=e^{-|x|/2}$ so that the associated $\rho$ (up to a normalization
constant) equals $e^{-|x|}$; hence $\rho$ is  just Lipschitz
continuous near the origin.

For the atomic case ($K=1; R_1=0, Z_1=Z$) a quantity studied earlier
is the spherical average of $\rho$ which, in polar coordinates 
$x=r\omega$ with $r=|x|$ and $\omega=x/|x|$, is defined by 
\begin{equation}\label{def:tilderho}
  \widetilde\rho(r)=\int_{\mathbb S^2}\rho(r\omega)\,d\omega\ , \quad
  r\in[0,\infty). 
\end{equation} 
The above mentioned analyticity result implies that $\widetilde \rho\in
C^{\omega}((0,\infty)).$ The existence of $\widetilde\rho\,'(0)$ and
the so-called
\textbf{cusp condition}  
\begin{equation}\label{eq:KatoCusp}
\widetilde \rho\,'(0)=-Z\widetilde\rho(0)
\end{equation}
follow from a similar result of Kato \cite{kato1} for $\psi$
itself; see also \cite{seiler} and \cite[Remark~1.13]{AHP}.
The existence of $\widetilde\rho\,''(0)$ and an implicit formula for
it was proved in \cite[Theorem 1.11]{AHP}; see \eqref{eq:secondAHP}
below for the exact statement. 

In \cite{CMP2} the present authors 
generalized the results of Kato for $\psi$ considerably 
for the Hamiltonian in \eqref{Hbis}.
 In the present paper we obtain results, partly in the
spirit of these findings, for the density $\rho$. In particular, we
prove results on the regularity of the density $\rho$ at the nuclei
and derive identities which the first and second radial derivatives of
\(\rho\) satisfy. 
These identities can be interpreted as cusp
conditions (analogously to \eqref{eq:KatoCusp}). The methods developed
in \cite{AHP} play an essential role in the proofs of these results. 

We indicate the importance of the electron density in quantum
mechanics. From the eigenfunction \(\psi\) it is, in principle,
possible to calculate the energy, various expectation values, etc.;
but \(\psi\) depends on \(3N\) variables. Physicists and chemists
usually aim at understanding atomic and molecular properties by means of
the electron density which is just a function on \(\R^3\) and can be
visualized. The density also has an immediate probabilistic
interpretation. 

In computational chemistry density functional methods are of
increasing importance for calculations of ground state energies of
large molecules. Thereby the energy is approximated by
minimizing a `density functional' which depends nonlinearly and
nonlocally upon the density. The minimizing function is believed to be
a good approximation to the density itself. The relationship
between most of these functionals and the full \(N\)-electron
Schr\"odinger equation  remains unclear though. One exception is of
course the archetype density functional 
theory, the Thomas-Fermi theory, which is mathematically
and physically interesting, and very well understood,
see \cite{LiebSimon} and  
\cite{Lieb1}.  For an interesting recent review on various 
mathematical problems related to the  many models in computational
chemistry,  see \cite{BrisLions}.
For some work on the density \(\rho\) from a numerical point of view,
related to regularity questions, see \cite{flad1}.

Questions concerning the one-electron density \(\rho\), as defined by
\eqref{rhohat}, pose some challenging
mathematical problems. Results as given in the present paper
contribute to a better understanding of the physics of atoms and
molecules and in addition should have relevance for computational
quantum chemistry.

In the following we use  the standard definition and notation for
H\"older continuity and  Lipschitz continuity, see e.g. \cite{GandT}. 
Let $f:\mathbb R^n \supset\Omega\to \mathbb R$, then $f\in
C^{k,\alpha}(\Omega)$ means, for 
$\alpha =0$, that $f$  is $k$ times continuously differentiable, for
$\alpha\in (0,1]$ that the $k$-th partial derivatives of $f$ are
H\"older continuous with exponent $\alpha$. In the case \(k=0\), we
often write \(C^{\alpha}(\Omega):=C^{0,\alpha}(\Omega)\) when
\(\alpha\in(0,1)\). 

The main result of the present paper is the following.
\begin{thm}\label{thm:ThmC-1-1}
Let $\psi\in L^2(\mathbb R^{3N})$ be a molecular or atomic N-electron
eigenfunction, i.e., \(\psi\) satisfies
\eqref{Hpsi}, with associated density $\rho$.
 Define $\mathcal F:\mathbb R^3\to \mathbb R$ by
\begin{equation}\label{def:F}
  \mathcal F(x)=-\sum_{k=1}^KZ_k|x-R_k|.
\end{equation}
Then 
\begin{equation}\label{mu}
  \rho(x)=e^{\mathcal{F}(x)}\mu(x) 
\end{equation}
with
\begin{equation}\label{muC11}
  \mu\in C^{1,1}(\mathbb R^3).
\end{equation}
This representation is optimal in the following sense: There is no 
function $\widetilde{\mathcal{F}}:\R^3\to\R$ depending only on
\(Z_1,\ldots,Z_K\), \(R_1,\ldots,R_K\), but neither on $N$, $\rho$, nor
\(E\), with the property that 
\(e^{-\widetilde{\mathcal{F}}}\rho\) is in \(C^2(\R^3)\).

Furthermore, $\mu$ admits the following representation:
\par\noindent There exist  
$C_1,\dots, C_K\in \mathbb R^3$ and $\nu:\mathbb
R^3\to \mathbb R$ such that 
\begin{equation}\label{munu}
  \mu(x)=\nu(x)+\sum_{k=1}^K|x-R_k|^2\big(
  C_k\cdot\frac{x-R_k}{|x-R_k|}\big), 
\end{equation}
with 
\begin{equation}\label{nualpha}
  \nu\in C^{2,\alpha}(\mathbb R^3) \text{ for all }\alpha\in (0,1).
\end{equation}
\end{thm}
\begin{remark}\label{at}
In the case of atoms (\(K=1; R_1=0, Z_1=Z\)), the statement of the
theorem reads: There exists 
\(C\in\R^3\) such that
\begin{equation}\label{eq:1}
  \rho(x)=e^{-Z|x|}\mu(x), \quad \mu(x)=\nu(x)+|x|^2\big(
  C\cdot\frac{x}{|x|}\big)
\end{equation} 
with
\begin{equation}\label{nuatom}
  \nu\in C^{2,\alpha}(\mathbb R^3) \text{ for all } \alpha\in (0,1).
\end{equation}
To simplify the exposition, we shall give the proof of
Theorem~\ref{thm:ThmC-1-1} only in the case of atoms.  
The proof easily generalizes to the case of several nuclei.
\end{remark}
\begin{remark}\label{thm:j-s}
It will be evident from the proof that the result (appropriately
reformulated) also holds for each
\(\rho_j\) seperately (see \eqref{rhohat}). The same is true for the
results below. 
\end{remark}
\begin{pf*}{Proof of the optimality}
We study `Hydrogenic atoms' ($N=K=1; R_1=0, Z_1=Z$) and use the
notation (contrary to the rest of the paper) $x = (x_1, 
x_2, x_3) \in {\mathbb R}^3$, $r=|x|$. In this case, the operator in
\eqref{Hbis} reduces to 
\(H_Z=-\Delta_x-Z/|x|\). 
We will present an example where, no matter what the choice of
\(\widetilde{\mathcal{F}}\) (as in the theorem), 
$\mu=e^{-\widetilde{\mathcal{F}}}\rho$ cannot be \(C^2\).
The argument 
resembles the proof of the corresponding result in \cite{CMP2}.

The $1s$ eigenfunction is \(\psi_{1s}(x)=e^{-Zr/2}\) with
\(H_Z\psi_{1s}=-(Z^2/4)\psi_{1s}\)
and the associated density is \(\rho_{1s}(x)=e^{-Zr}\). 
The $2s$ and $2p$ eigenfunctions are
\begin{align*}
  \psi_{2s}(x) &= (1-\tfrac{Z}{4}r)e^{-Zr/4} , &
  \psi_{2p}(x) &= x_1 e^{-Zr/4}.
\end{align*}
Both satisfy \(H_Z\psi=E\psi\) with $E=-Z^2/16$.
The associated densities are
\begin{align*}
  \rho_{2s}(x) &= \psi_{2s}^2(x) = (1-\tfrac{Z}{4}r)^2e^{-Zr/2}, &
  \rho_{2p}(x) &= \psi_{2p}^2(x) = x_1^2 e^{-Zr/2}.
\end{align*}
Consider now $\psi_{{\rm mixed}}= \psi_{2s} + \psi_{2p}$ and
\begin{align*}
  \rho_{{\rm mixed}} = \psi_{{\rm mixed}}^2 =
  \rho_{2s}+ \rho_{2p} + 2 \psi_{2s} \psi_{2p}.
\end{align*}
A simple calculation shows that
\begin{align*}
  e^{Zr} \rho_{2s}, e^{Zr} \rho_{2p} \in C^{2,1}({\mathbb R}^3),
\end{align*}
but \(e^{Zr}\rho_{\rm mixed}\) is just \(C^{1,1}\),  since the mixed 
derivative \(\partial_{x_2}\partial_{x_1}\) of 
\begin{align*}
  e^{Zr} \psi_{2s} \psi_{2p} = x_1 e^{Zr/2}(1-\tfrac{Z}{4})
\end{align*}
does not exist at \(x=0\). 

But if \(\rho=e^{\widetilde{\mathcal{F}}}\mu\) with \(\mu\in C^2\), then
\begin{align*}
  \frac{\mu_{\rm
      mixed}}{\mu_{1s}}=\frac{e^{-\widetilde{\mathcal{F}}}\rho_{\rm
      mixed}}{e^{-\widetilde{\mathcal{F}}}e^{-Zr}}
    =e^{Zr}\rho_{\rm mixed}
\end{align*}
should also be \(C^2\), a contradiction.
\end{pf*}

Note that  $\psi_{2s}(x)=\psi_{2s}(-x)$ and
$\psi_{2p}(x)=-\psi_{2p}(-x)$, but their linear combination  
$\psi$ is neither even nor odd.
\begin{remark}
The representation  of $\rho$ as a product  $\rho=e^{\mathcal F}\mu$
with a fixed 
`universal' $\mathcal F$ such that $\mu$ is by one degree smoother
than $\rho$ corresponds to Theorem 1.1 in \cite{CMP2} where a similar
result was obtained for the eigenfunction \(\psi\) itself. In that
case though, the correponding $\mathcal F$ is more 
complicated since many-particle interactions have to be taken into
account. For some interesting recent investigation in connection with
Jastrow factors from a numerical point of view, see \cite{flad2}.
\end{remark}
The proof of Theorem \ref{thm:ThmC-1-1} will be given in the next
section. Here we just mention that $\rho$ satisfies an inhomogeneous
Schr\"odinger equation 
whose investigation is crucial for regularity results like the above,
as well as it was for the results  in \cite{AHP}. 
Let $H$ be given by \eqref{Hbis} and consider an eigenfunction $\psi$
satisfying \eqref{Hpsi}. To simplify notation we assume without loss
that \(\psi\) is real. The equation
\begin{equation}\label{hatint}
  \int_{\mathbb R^{3N-3}}\psi(x, \mathbf{\hat
    x}_j)(H-E)\psi(x,\mathbf{\hat x}_j)\,d 
  \mathbf{\hat x}_j=0
\end{equation}   
leads to an equation (in the sense of distributions) for $\rho_j$,
namely, 
\begin{equation}
 \label{eq:first-rho}
 \Big(-\frac{1}{2}\Delta
 -\sum_{k=1}^K\frac{Z_k}{|x-R_k|}\Big)\rho_j+h_j=0. 
\end{equation}
Summing \eqref{eq:first-rho} over $j$ we obtain
the equation for $\rho$,
\begin{equation}\label{eq:firstRho}
  \Big(-\frac{1}{2}\Delta -\sum_{k=1}^K\frac{Z_k}{|x-R_k|}\Big)\rho+h=0,
\end{equation}
with $h=\sum_{j=1}^Nh_j$.
The functions $h_j$ will be given explicitely in
Section~\ref{sec:proofs}; see \eqref{eq:defh}.

In \cite{AHP} we considered the spherically averaged density
\(\widetilde\rho\) (as defined by
\eqref{def:tilderho}) for the atomic
case. The regularity of \(\widetilde h\) (the spherical average of
\(h\) above) was crucial for the results obtained there. Here we study
the {\it non-averaged} density \(\rho\) for the general case of
molecules. Again, the regularity of \(h\) is essential for our
results. 

We continue to consider $\rho$ in the neighbourhood of one nucleus
with charge $Z$. Without loss we can place this nucleus at the origin.

The equations \eqref{munu} and \eqref{nualpha} show that it is natural
to consider the behaviour  of $\rho(r\omega)$ for fixed $\omega$ as
$r$ tends to zero.  

\begin{thm}\label{thm:dirDerI}
Let $\psi\in L^2(\mathbb R^{3N})$ be a molecular or atomic
eigenfunction, i.e., \(\psi\)
satisfies \eqref{Hpsi}, with
associated density $\rho$. Assume without loss that 
\(R_1=0\) and write
$Z$ instead of \(Z_1\). 
Let $r_0=\min_{k>1}|R_k|$ (\(r_0=\infty\) for atoms) and let
\(\omega\in\mathbb{S}^2\) be fixed. 
\begin{itemize}
\item[(i)] 
The function $r\mapsto \rho(r,\omega):=\rho(r\omega)$, $r\in [0,r_0)$,
satisfies 
\begin{align}\label{eq:rhoOmega}
  \rho(\cdot, \omega)\in C^{2,\alpha}([0,r_0))\text{ for all }\alpha\in
  (0,1).
\end{align}
\item[(ii)] Denote by \({}'\) the derivative \(\frac{d}{dr}\), and
  define 
\begin{align}
  \eta(x)=e^{Z|x|}\rho(x)\ , \ 
  \chi =\eta - r^2( C\cdot\omega),
\end{align}
where \(C\in\R^3\) is the constant \(C_1\) in \eqref{munu} (resp.\
\(C\) in \eqref{eq:1}).

Then
\begin{align}\label{reg:eta&chi}
\eta\in C^{1,1}(B(0,r_0))\ , \ 
\chi\in C^{2,\alpha}(B(0,r_0))\text{ for all }\alpha\in(0,1),
\end{align}
and
\begin{align}\label{eq:fixedW3}
  \rho'(0,\omega)&
   =-Z\rho(0)+
  \omega\cdot (\nabla \eta)(0),
  \\
  \label{eq:fixedW4}
  \rho''(0,\omega)
  &=Z^2\rho(0)+2\omega\cdot[C-Z(\nabla \eta)(0)]+
  \omega\cdot\big((D^2\chi)(0)\omega\big).
\end{align}
Here $(D^2\chi)(0)$ is the Hessian matrix of
$\chi$ evaluated at the origin.  
\end{itemize}
\end{thm}

\begin{remark}\label{rem:new cusps}
  \(\, \)
\begin{enumerate}
\item[\rm (i)] For atoms, \(\eta\) equals \(\mu\) from
  Theorem~\ref{thm:ThmC-1-1} and \(\chi\) equals \(\nu\) from
  Remark~\ref{at}. 
\item[\rm (ii)]
Note that \eqref{eq:rhoOmega} trivially implies that
\(\rho(r,\omega)=e^{-Zr}\eta(r,\omega)\) with \(\eta(\cdot,\omega)
\in C^{2,\alpha}([0,r_0))\) for all \(\alpha\in(0,1)\). Compare with
\eqref{mu}, \eqref{muC11}.
\item[\rm (iii)] In \cite[Theorem~1.11]{AHP} it was proved that
  \(\widetilde \rho\) defined by \eqref{def:tilderho} belongs to
  \(C^2([0,r_0))\cap C^{2,\alpha}((0,r_0))\) for all
  \(\alpha\in(0,1)\). (The proof in \cite{AHP} for the atomic case
  easily generalizes to the molecular case.)
  Reading the proof of \cite[Theorem~1.11]{AHP} carefully, one
  sees that 
  it in fact yields \(\widetilde \rho\in C^{2,\alpha}([0,r_0))\).
 The statement in \eqref{eq:rhoOmega} shows that for fixed
 \(\omega\in\mathbb{S}^2\) this holds already for
 \(\rho(\cdot,\omega)\), i.e.,  without averaging. 
\item[\rm (iv)]
The identities \eqref{eq:fixedW3} and \eqref{eq:fixedW4} can be
considered as \textbf{non-isotropic 
cusp conditions of first and second order}. They generalize the cusp
condition \eqref{eq:KatoCusp}, as well as the previously mentioned 
result in \cite{AHP} for \(\widetilde\rho\,''(0)\); more on this in
Remark~\ref{mu21} (ii) below. See also the second order cusp 
conditions obtained in \cite{CMP2} for the eigenfunction \(\psi\)
itself. 
\item[\rm (v)]
It is worth noting that \eqref{eq:fixedW3} and \eqref{eq:fixedW4}
can be interpreted  
as a structural result for the density $\rho$: From
Theorem~\ref{thm:dirDerI} it follows
that in a neighbourhood of a nucleus (which is at the origin), $\rho$
satisfies (for all \(\alpha\in(0,1)\))
\begin{equation}\label{formalrho}
  \rho(r,\omega)=\rho(0)+r\phi_1(\omega)+r^2\phi_2(\omega)+O(r^{2+\alpha})\
  ,\ r\downarrow0,
\end{equation}
and \eqref{eq:fixedW3}, \eqref{eq:fixedW4} show that $\phi_1$ is a
linear and $\phi_2$ a quadratic 
polynomial restricted to $\mathbb S^2$.\par\noindent
It is a natural question whether 
\eqref{formalrho} extends to higher orders.  
\end{enumerate}
\end{remark} 
We continue with the atomic case. In view of Remark~\ref{at}, \eqref{eq:1} 
and the considerations after the proof of the optimal regularity of
\(\mu\) in Theorem \ref{thm:ThmC-1-1},  
the following theorem is natural.   
 
\begin{thm}\label{thm:ThmC-2-alpha}
Let $\psi\in L^2(\R^{3N})$ be an atomic eigenfunction
with associated density $\rho$. 
Suppose that 
\begin{equation}\label{symmetry}
  |\psi(\mathbf x)|=|\psi(-\mathbf x)|\text{ for all }\mathbf x\in
  \mathbb R^{3N}. 
\end{equation}
Then $\rho$ satisfies 
\begin{equation}\label{main3}
  \rho(x)=e^{-Z|x|}\mu(x),\:\: \mu\in C^{2,\alpha}(\mathbb R^3)\text{
  for all } \alpha \in (0,1). 
\end{equation} 
Furthermore, 
\begin{equation}\label{cusp0}
  \rho'(0,\omega)=-Z\rho(0)\ , \quad
  \rho''(0,\omega)=Z^2\rho(0)+
  \omega\cdot\big((D^2\mu)(0)\omega\big).
\end{equation}
We also have
\begin{align}\label{eq:Marias}
   \rho''(0,\omega)=\frac23\big(
   Z^2\rho(0)+h(0,\omega)\big)
   +\frac13\lim_{r\downarrow0}\frac{(\mathcal{L}^2\rho)(r,\omega)}{r^2}, 
\end{align}
with \(h\) from \eqref{eq:firstRho}, and \(\mathcal{L}^2/r^2\) the
angular part of 
\(-\Delta\), i.e., \(\Delta=\partial^2/\partial
r^2+(2/r)\partial/\partial r-\mathcal{L}^2/r^2\). 
\end{thm}
\begin{remark}\label{mu21}
  \(\, \)
\begin{enumerate}
\item[\rm (i)] 
In this case \(\mu=\nu=\chi=\eta\) , as can be seen from
Remark~\ref{rem:new cusps} (i)  and the proof of the theorem.
\item[\rm (ii)] Note that \eqref{cusp0} shows that the cusp
  condition \eqref{eq:KatoCusp} in this case holds for fixed angle
  \(\omega\in\mathbb{S}^2\) without averaging. 
 Further, taking the
  spherical average of \eqref{eq:Marias}, we get the formula for
  \(\widetilde\rho\,''(0)\) obtained in \cite[Theorem~1.11 (iv)]{AHP}:
  \begin{align}\label{eq:secondAHP}
     \widetilde\rho\,''(0)=\frac23\big(Z^2\widetilde\rho(0)+\widetilde
     h(0)\big). 
  \end{align}
  To see this note that for all \(r>0\)
  \begin{align*}
  \int_{\mathbb{S}^2}1\cdot({\mathcal{L}}^2\rho)(r,\omega)\,d\omega
  =\int_{\mathbb{S}^2}(\mathcal{L}^21)\cdot\rho(r,\omega)\,d\omega=0.
  \end{align*}
  Note that \(\widetilde\rho\,''(0)\geq0\), since 
  \begin{align*}
    \widetilde h(r)\geq \epsilon\widetilde\rho(r)
  \end{align*}
  for some \(\epsilon\geq0\)
  \cite[Theorem 1.11]{AHP}. This
  positivity is not an obvious consequence of the formula
  in \eqref{cusp0}.
  \item[\rm (iii)] As can be seen from the proof of
    Theorem~\ref{thm:ThmC-2-alpha}, \(h\in C^{\alpha}(\R^3)\) for all
    \(\alpha\in(0,1)\) in this case.
\end{enumerate}
\end{remark}

\section{Proofs}
\label{sec:proofs}
\begin{pf*}{Proof of Theorem~\ref{thm:ThmC-1-1}}
As noted in Remark~\ref{at}, we shall give the proof
only in the case of atoms (\(K=1; R_1=0, Z_1=Z\)). 

For the regularity questions concerning $\rho$ defined in \eqref{rhohat}
it suffices to consider the 
(non-symmetrized) density $\rho_1$ defined by
\begin{align}
  \label{eq:rho}\nonumber
  \rho_1(x)&= \int_{{\mathbb R}^{3N-3}}
  \,|\psi(x,x_2,\ldots,x_N)|^2\,dx_2\cdots dx_N
  \\&
  =\int_{\mathbb R^{3N-3}}|\psi(x,\hat{\bf x}_1)|^2\,d\hat{\bf x}_1
\end{align}
with \(x\in\mathbb R^3, \hat{\bf
  x}_1=(x_2,\ldots,x_N)\in\mathbb R^{3N-3}\). 

As explained in \eqref{hatint}--\eqref{eq:firstRho} \(\rho_1\)
satisfies the  Schr\"{o}dinger-type equation 
\begin{align}\label{eq2:rho}
  {}-\Delta \rho_1 - \frac{2Z}{|x|}\rho_1  + 2h_1=0, 
\end{align}
where the function $h_1$ is given by 
\begin{align}
\label{eq:defh}
  h_1(x) &= J_1 - J_2 + J_3 - E\rho_1(x),\\ 
  J_1(x) &= \sum_{j=1}^N \int _{{\mathbb R}^{3N-3}} |\nabla_j \psi |^2
  \,d\hat{\bf x}_1\ ,\quad
  J_2(x) = \sum_{j=2}^N \int _{{\mathbb R}^{3N-3}} \frac{Z}{|x_j|} \psi^2
  \,d\hat{\bf x}_1,\nonumber\\ 
  J_3(x) &= \sum_{k=2}^N \int _{{\mathbb R}^{3N-3}} \frac{1}{|x-x_k|}
  \psi ^2 \,d\hat{\bf x}_1
  +\sum_{2\leq j<k\leq N} \int _{{\mathbb R}^{3N-3}} \frac{1}{|x_j-x_k|}
  \psi ^2 \,d\hat{\bf x}_1.
  \nonumber
\end{align}
(We will henceforth partly omit the variables in the integrands).
Using the exponential decay of \(\psi\) \eqref{eq:exp-dec} and of
\(\nabla\psi\) \eqref{eq:dec_grad_psi} 
one can prove that  \(h_1\in L^{\infty}(\mathbb R^3)\) (for details,
see \cite[Theorem 1.11]{AHP}).

Making the {\it Ansatz} 
\begin{align}
  \label{eq:ansatzRho}
  \rho_1(x) = e^{-Z|x|}\mu_1(x)
\end{align}
and using \eqref{eq2:rho},  
we get that $\mu_1$ satisfies the  equation 
\begin{align}\label{eq:mu}
  \Delta \mu_1 
  = 2e^{Z|x|} h_1 + 2Z
  \omega\cdot\nabla \mu_1 - Z^2\mu_1.
\end{align}
Here $\omega = \frac{x}{|x|}$. Since \(\rho_1\in C^{0,1}(\mathbb
R^3)\) (as mentioned in the introduction), also \(\mu_1\in C^{0,1}(\mathbb
R^3)\).
Clearly, the function $x \mapsto
\omega$ belongs to \(L^{\infty}(\mathbb R^3)\).
The fact that also $x \mapsto h_1(x)$ 
is in \(L^{\infty}(\mathbb R^3)\)
gives, by standard elliptic regularity
\cite[Theorem 10.2]{LiebLoss}, that 
\begin{align}\label{firstMu}
  \mu_1 \in
  C^{1,\alpha}(\mathbb R^3) \text{ for all } \alpha\in(0,1).
\end{align}
Our aim is to prove more, namely that
\begin{align}
 \label{eq:muNew1} 
 \Delta\mu_1=c_1\cdot\omega+g\,,\ c_1\in\mathbb R^3\, ,\ g\in
  C^{\alpha}(\mathbb R^3) \text{ for all } \alpha\in(0,1).
\end{align}

Since, by \eqref{firstMu}, \(\nabla\mu_1\) is continuous
at the origin, the term \(2Z \omega\cdot \nabla \mu_1\) behaves like
\(c_1^{(1)}\cdot\omega\) (\(c_1^{(1)}=2Z\nabla\mu_1(0)\in\mathbb
R^3\)) at the origin. 
It turns out that generally
$h_1$ is discontinuous at the origin, also behaving like $c_1^{(2)}\cdot
\omega$ (\(c_1^{(2)}\in\mathbb R^3\)).
However, one can solve the equation
\(\Delta u_1=c_1\cdot\omega\) (\(c_1\in\mathbb R^3\))
explicitely, and one gets that the solution \(u_1\) is \(C^{1,1}\). From 
standard elliptic regularity, the other terms give contributions which
belong to \(C^{2,\alpha}(\mathbb R^3)\). Below we give the details.

First, we write
\begin{align}
  \label{eq:firstOrder}
  2Z\omega\cdot \nabla \mu_1(x)
  &=c_1^{(1)}\cdot\omega+ g_1(x)\ , \ g_1\in C^{\alpha}(\mathbb
  R^3)\,,\,\alpha\in(0,1), \\
  c_1^{(1)}&=2Z\nabla\mu_1(0)\ , \
  g_1(x)=2Z\omega\cdot\big(\nabla\mu_1(x)-\nabla\mu_1(0)\big). 
  \nonumber
\end{align}
That \(g_1\in C^{\alpha}(\mathbb R^3)\), \(\alpha\in(0,1)\), follows from
Lemma~\ref{usefullemma} in Appendix~\ref{app:lem}.

We next consider $h_1$ as defined in \eqref{eq:defh}. We will show the
following: 
\begin{lemma}\label{lem:formH}
Let \(h_1\) be as in \eqref{eq:defh}. Then there exist \(c_1^{(2)}\in
\R^3\), \(G:\R^3\to\R\), such that
\begin{align}
  \label{eq:claim}
  h_1=c_1^{(2)}\cdot\omega+G\,,\ G\in C^{\alpha}(\mathbb R^3)\,,\,
  \alpha\in(0,1). 
\end{align}
\end{lemma}
Before proving Lemma~\ref{lem:formH}, we finish the proof of
Theorem~\ref{thm:ThmC-1-1}. 

Lemma~\ref{lem:formH} and Lemma~\ref{usefullemma} in
Appendix~\ref{app:lem} imply 
that 
\begin{align*}
  (e^{Z|x|}-1)h_1\in   C^{\alpha}(\mathbb R^3)\quad\text{ for all }
  \alpha\in(0,1),
\end{align*}
and it therefore follows 
from \eqref{eq:mu}, \eqref{firstMu}, \eqref{eq:firstOrder}, and
\eqref{eq:claim} that
\begin{align}
  \label{eq:muNew2}
  \Delta\mu_1= c_1\cdot\omega+ g\ , \qquad
  c_1&=c_1^{(1)}+2c_1^{(2)}\in\mathbb R^3\ ,\\g&\in
  C^\alpha(\mathbb R^3)\,,\, 
  \alpha\in(0,1).\nonumber
\end{align}
This is \eqref{eq:muNew1}, which we aimed to prove.

A simple computation shows that the function \(u_1(x)=\frac16|x|^2
c_1\cdot\omega=\frac{|x|}{6}\,c_1\cdot x\) satisfies \(\Delta
u_1=c_1\cdot\omega\), and so \(\nu_1=\mu_1-u_1\) solves \(\Delta \nu_1=g\), \(g\in
C^{\alpha}(\mathbb R^3)\), \(\alpha\in(0,1)\). From standard elliptic
regularity theory \cite[Theorem 10.3]{LiebLoss} follows that \(\nu_1\in
C^{2,\alpha}(\mathbb R^3)\), \(\alpha\in(0,1)\). Note that due to
Lemma~\ref{usefullemma}, \(u_1\in C^{1,1}(\mathbb R^3)\), and so
\begin{align}
\label{eq:mu=v+u}
\mu_1=\nu_1+u_1=\nu_1+\frac16|x|^2(
c_1\cdot\omega)\in C^{1,1}(\mathbb R^3).
\end{align}
This finishes the proof of Theorem~\ref{thm:ThmC-1-1} for atoms, with
\begin{align}\label{def:bigC}
  C=\frac16\sum_{j=1}^Nc_j,
\end{align}
where \(c_j\) is the contribution from
\(\rho_j\). 
\end{pf*}
It remains to prove Lemma~\ref{lem:formH}.
\begin{pf*}{Proof of Lemma~\ref{lem:formH}}
The proof is essentially a tedious but elementary verification, the
idea being to isolate and extract the most singular term of
\(h_1\). Part of this has been carried out in \cite{AHP}, and, in
order not to repeat the details, we refer to that paper whenever
possible. We also use the same notation.

Define
\begin{align}
  \label{eq:psi1}
    \psi_1 &= e^{-(F -F_1)} \psi,
\end{align}
with
\begin{align}
  \label{def:Fatom}
  F({\bf x})&= -\frac{Z}{2}\sum_{j=1}^N |x_j| + \frac{1}{4}
  \sum_{1\leq j<k\leq N} |x_j -x_k|
  ,\\
  \label{def:F1} 
  F_1({\bf x})&= -\frac{Z}{2}\sum_{j=1}^N \sqrt{|x_j|^2+1} + \frac{1}{4}
  \sum_{1\leq j<k\leq N} \sqrt{|x_j -x_k|^2+1}.
\end{align}
Then it follows from \cite[Proposition 1.5]{AHP} that $\psi_1 \in
C^{1,\alpha}(\mathbb R^3), \alpha\in(0,1)$.

In  \cite[Lemma 3.5 (i)]{AHP} it is proven that (with the notation from
\eqref{eq:defh}) $J_2, J_3 \in C^{\alpha}({\mathbb
  R}^3), \alpha\in(0,1)$. Furthermore, using \eqref{eq:psi1},  
$J_1$ is written as
\begin{align*}
  &J_1(x)  = \int_{{\mathbb R}^{3N-3}} |\nabla\psi|^2 \,d\hat{\bf x}_1
  = I_1 + I_2 + I_3 + I_4 + I_5 + I_6, 
\end{align*}
where
\begin{align*}
 I_1(x) &= \sum_{j=1}^N \int _{{\mathbb R}^{3N-3}} |\nabla_j F|^2 \psi^2
  \,d\hat{\bf x}_1,\nonumber\\ 
 I_2(x)&=  \sum_{j=1}^N\int_{{\mathbb
      R}^{3N-3}}|\nabla_j F_1|^2\psi^2 \,d\hat{\bf x}_1,\nonumber
  \\ 
 I_3(x) &= -2 \sum_{j=1}^N \int _{{\mathbb R}^{3N-3}} (\nabla_j F\cdot
  \nabla_j F_1) \psi^2 \,d\hat{\bf x}_1,\nonumber\\
  I_4(x)&= \sum_{j=1}^N\int_{{\mathbb R}^{3N-3}} 
  e^{2(F-F_1)}|\nabla_j\psi_1|^2\,d\hat{\bf x}_1,\nonumber
  \\
 I_5(x) &= 2\sum_{j=1}^N \int _{{\mathbb R}^{3N-3}}  (\nabla_j F\cdot
  \nabla_j \psi_1) e^{2(F-F_1)} \psi_1 \,d\hat{\bf x}_1, \nonumber\\
  I_6(x)&= -2\sum_{j=1}^N
  \int_{{\mathbb
      R}^{3N-3}}
  (\nabla_j 
  F_1\cdot \nabla_j\psi_1)e^{2(F-F_1)}\psi_1\,d\hat{\bf x}_1.
\end{align*}
It is proven in \cite[p. 93, bottom]{AHP} that
\begin{align*}
  I_2, & I_4, I_6 \in C^{\alpha}({\mathbb R}^3)\ , \ 
  \alpha\in(0,1).\nonumber
\end{align*}
It is also proven in \cite{AHP} that 
\begin{align}
  \label{eq:Is}
  I_1(x) &= \int_{{\mathbb R}^{3N-3}} |\nabla_1 F|^2 \psi^2 \,d\hat{\bf x}_1 +
  \tilde{I}_1(x),\nonumber\\ 
  I_3(x) &= -2 \int _{{\mathbb R}^{3N-3}} (\nabla_1 F\cdot \nabla_1 F_1)
  \psi^2 \,d\hat{\bf x}_1 + \tilde{I}_3(x),\nonumber\\ 
  I_5(x) &= 2\int _{{\mathbb R}^{3N-3}}  (\nabla_1 F\cdot \nabla_1 \psi_1)
  e^{2(F-F_1)} \psi_1 \,d\hat{\bf x}_1 + \tilde{I}_5(x), 
\end{align}
with $\tilde{I}_j \in C^{\alpha}(\mathbb R^3), \alpha\in(0,1)$. (For
\(\tilde I_1\), this is \cite[(3.52) and until (3.53)]{AHP};
for \(\tilde I_3\) and \(\tilde I_5\), this is \cite[(3.55) and
(3.56), and between]{AHP}).

Now, (see \eqref{def:Fatom}) 
\begin{align*}
  |\nabla_1 F|^2(x,\hat{\bf x}_1) &=
  \frac{Z^2}{4} - \frac{Z}{4} \frac{x}{|x|}\cdot \sum_{j=2}^N
  \frac{x-x_j}{|x-x_j|} 
  +\frac{1}{16} \Big(\sum_{j=2}^N \frac{x-x_j}{|x-x_j|} \Big)^2. 
\end{align*}
Therefore, we write
\begin{align*}
  &\int _{{\mathbb R}^{3N-3}} |\nabla_1 F|^2 \psi^2 \,d\hat{\bf x}_1
  = I_{1,1}(x) + I_{1,2}(x), \nonumber\\
  &I_{1,1}(x)=
  - \frac{Z}{4} \frac{x}{|x|}\cdot \int _{{\mathbb R}^{3N-3}}
  \sum_{j=2}^N \frac{x-x_j}{|x-x_j|} 
  \psi^2 \,d\hat{\bf x}_1 ,\nonumber\\
  &I_{1,2}(x)= \int _{{\mathbb R}^{3N-3}}
  \Big\{\frac{Z^2}{4} +
  \frac{1}{16} \Big(\sum_{j=2}^N \frac{x-x_j}{|x-x_j|} \Big)^2\Big\}
  \psi^2 \,d\hat{\bf x}_1. 
\end{align*}
The proof that $I_{1,2}$ belongs to
$C^{\alpha}(\R^3)$, \(\alpha\in(0,1)\),  is also in \cite{AHP} (between
(3.52) and (3.55)).

For $I_{1,1}$, we write 
$I_{1,1}(x) = \frac{Z}{4} \frac{x}{|x|}\cdot {\rm
  Int}_1(x)$, with 
\begin{align}\label{int1}
  {\rm Int}_1(x) = -\int _{{\mathbb R}^{3N-3}} \sum_{j=2}^N
  \frac{x-x_j}{|x-x_j|} 
  \psi^2\,d\hat{\bf x}_1.
\end{align}
The function ${\rm Int}_1$ 
belongs to $C^{\alpha}(\mathbb R^3)$, \(\alpha\in(0,1)\).
This follows by arguments as for the integral
\begin{align*}
  \int\frac{1}{|x-x_k|}\big(x\cdot(x-x_k)\psi^2\big)
  \,dx_2\cdots dx_N
\end{align*}
in \cite{AHP} (see between (3.54) and (3.55)).

Therefore (with \(\omega=\frac{x}{|x|}\)),
\begin{align}\label{omegaInt1}
  I_{1,1}(x)&=c_{1,1}\cdot\omega+g_{1,1}(x)\ , \ g_{1,1}\in
  C^{\alpha}(\mathbb R^3)\,,\,\alpha\in(0,1),\\
  \label{omegaInt1Bis}
  c_{1,1}&=\frac{Z}{4}{\rm Int}_1(0)\ , \
  g_{1,1}(x)=\frac{Z}{4}\omega\cdot\big({\rm 
  Int}_1(x)- {\rm
  Int}_1(0)\big).
\end{align}
That \(g_{1,1}\in C^{\alpha}(\mathbb R^3)\), \(\alpha\in(0,1)\),
follows from Lemma~\ref{usefullemma} in Appendix~\ref{app:lem}.

Next we consider $I_3$ (see \eqref{eq:Is}). 
Since
\begin{align}
  \label{eq:nablaF}
  \nabla_1 F(x,\hat{\bf x}_1) &=
  -\frac{Z}{2} \frac{x}{|x|} + \frac{1}{4} \sum_{j=2}^N \frac{x -
  x_j}{|x-x_j|}, \\
  \nabla_1 F_1(x,\hat{\bf x}_1) &=
  -\frac{Z}{2} \frac{x}{\sqrt{|x|^2+1}} + \frac{1}{4} \sum_{j=2}^N
  \frac{x - x_j}{\sqrt{|x-x_j|^2+1}}, \nonumber
\end{align}
we have
\begin{align*}
  &\int _{{\mathbb R}^{3N-3}} (\nabla_1 F\cdot \nabla_1 F_1) \psi^2
  \,d\hat{\bf x}_1 
  = I_{3,1}(x) + I_{3,2}(x), \nonumber\\
  &I_{3,1}(x) = -\frac{Z}{8} \sum_{j=2}^N \frac{x}{|x|}\cdot
  \int_{\R^{3N-3}}  
  \frac{x-x_j}{\sqrt{|x-x_j|^2+1}} \psi^2 d\hat{\bf x}_1,\nonumber\\ 
  &I_{3,2}(x) = \int_{\R^{3N-3}}
  \Big\{\frac{Z^2|x|}{4\sqrt{|x|^2+1}}+\big(\frac{1}{4} \sum_{j=2}^N
  \frac{x - x_j}{|x-x_j|} \big)\cdot \nabla_1 F_1
  \Big\}\psi^2\,d\hat{\bf x}_1. 
\end{align*}
That the first term in \(I_{3,2}\) belongs to $C^{\alpha}(\R^3)$,
\(\alpha\in(0,1)\) follows by arguments as in \cite{AHP} (by 
applying Lemma~3.4 as done after (3.50); note that the function
\(x\mapsto |x|/\sqrt{|x|^2+1}\) belongs to \(C^{0,1}(\R^3)\)). 

That the last term in \(I_{3,2}\) belongs to $C^{\alpha}(\R^3)$,
\(\alpha\in(0,1)\), is proved in
\cite[(3.55), and after]{AHP}.

For \(I_{3,1}\), we write \(I_{3,1}(x)=-\frac{Z}{8}\frac{x}{|x|}\cdot
{\rm Int}_3(x)\), with 
\begin{align}\label{int3}
  {\rm Int}_3(x) = \sum_{j=2}^N \int_{\R^{3N-3}}  \frac{x -
  x_j}{\sqrt{|x-x_j|^2+1}} \psi^2 d\hat{\bf x}_1.
\end{align}
Similar arguments as for the first integral in  \(I_{3,2}\) above show that
\({\rm Int}_3\)
belongs to $C^{\alpha}(\R^3)$, \(\alpha\in(0,1)\). 

This implies, by Lemma~\ref{usefullemma}, that
\begin{align}\label{omegaInt2}
   I_{3,1}(x)&=c_{3,1}\cdot\omega+g_{3,1}(x)\ , \ g_{3,1}\in
  C^{\alpha}(\mathbb R^3)\,,\,\alpha\in(0,1),\\
  \label{omegaInt2Bis}
  c_{3,1}&=-\frac{Z}{8}{\rm Int}_3(0)\ , \
  g_{3,1}(x)=-\frac{Z}{8}\omega\cdot\big({\rm 
  Int}_3(x)- {\rm
  Int}_3(0)\big).
\end{align}

Finally, we consider $I_5$ (see \eqref{eq:Is}).
We use the same kind of analysis.
Using \eqref{eq:nablaF}, we write
\begin{align*}
  &\int_{{\mathbb R}^{3N-3}}  (\nabla_1 F\cdot \nabla_1 \psi_1)
  e^{2(F-F_1)} \psi_1 \,d\hat{\bf x}_1  
  = I_{5,1}(x) + I_{5,2}(x), \nonumber\\
  & I_{5,1}(x) = -\frac{Z}{4} \frac{x}{|x|}\cdot \int_{\R^{3N-3}}  e^{2(F-F_1)}\nabla_1
  (\psi_1^2) \,d\hat{\bf x}_1,\nonumber\\ 
  & I_{5,2}(x) =  \frac{1}{4} \sum_{j=2}^N \int_{\R^{3N-3}} \frac{x - x_j}{|x-x_j|}
  \cdot(\nabla_1\psi_1) e^{2(F-F_1)}\psi_1 \,d\hat{\bf x}_1. 
\end{align*}
Again, the proof that $I_{5,2}$ belongs to
$C^{\alpha}(\R^3)$, \(\alpha\in(0,1)\), is in 
\cite[(3.56), and after]{AHP}.
As before, we write \(I_{5,1}(x)=-\frac{Z}{4}\frac{x}{|x|}\cdot {\rm
  Int}_5(x)\) with 
\begin{align}\label{int5}
  {\rm Int}_5(x) =  \int_{\R^{3N-3}}  e^{2(F-F_1)}\nabla_1
  (\psi_1^2) \,d\hat{\bf x}_1.
\end{align}
The  integral \({\rm Int}_5\) belongs to $C^{\alpha}(\R^3)$,
\(\alpha\in(0,1)\). This follows by arguments as in \cite{AHP} (the
term in (3.56) with \(j=1\)). Therefore, by Lemma~\ref{usefullemma},
 \begin{align}
   \label{eq:I-5-1}
     I_{5,1}(x)&=c_{5,1}\cdot\omega+g_{5,1}(x)\ , \ g_{5,1}\in
  C^{\alpha}(\mathbb R^3)\,,\,\alpha\in(0,1),\\
  \label{eq:c-5-1}
  c_{5,1}&=-\frac{Z}{4}{\rm Int}_5(0)\ , \
  g_{5,1}(x)=-\frac{Z}{4}\omega\cdot\big({\rm 
  Int}_5(x)- {\rm
  Int}_5(0)\big).
 \end{align}

It follows from all of the above that
\begin{align}
  \label{eq:hFinal}
   h_1&=(c_{1,1}-2c_{3,1}+2c_{5,1})\cdot\omega
   +\big[(g_{1,1}-2g_{3,1}+2g_{5,1})
   \\&\qquad\qquad
   +(I_{1,2}-2I_{3,2}+2I_{5,2})+(I_2+I_4+I_6)
   \nonumber\\&\qquad\qquad\qquad\qquad\qquad
   +(\tilde I_1+\tilde
   I_3+\tilde I_5)-J_2+J_3\big]-E\rho_1
  \nonumber\\&
   =(c_{1,1}-2c_{3,1}+2c_{5,1})\cdot\omega + f - E\rho_1, \text{ with } f\in
   C^{\alpha}(\mathbb R^3)\ ,\  \alpha\in(0,1).
  \nonumber
\end{align}
Since \(\rho_1\in C^{0,1}(\R^3)\),
\eqref{eq:hFinal}
shows that \(h_1\) indeed can be written as in
\eqref{eq:claim}; that is, this 
finishes the proof of 
Lemma~\ref{lem:formH}.
\end{pf*}
\begin{pf*}{Proof of Theorem~\ref{thm:dirDerI}}
That \(\eta\in  C^{1,1}(B(0,r_0))\) follows from
\eqref{def:F}--\eqref{muC11}. 
That \(\chi\in C^{2,\alpha}(B(0,r_0)), \alpha\in(0,1)\), is a consequence of
\eqref{munu}--\eqref{nualpha}. 
It remains to prove \eqref{eq:rhoOmega}, \eqref{eq:fixedW3},
\eqref{eq:fixedW4}.  

As for Theorem~\ref{thm:ThmC-1-1}, 
we shall only give the proof for the case of atoms (\(K=1; R_1=0,
Z_1=Z, \eta=\mu\) and \(\chi=\nu\)).  

Let in the sequel \(\omega=\frac{x}{|x|}\in\mathbb{S}^2\) be
arbitrary, but fixed.  
Note that \eqref{eq:1} and \eqref{nuatom}
imply that
\(\rho(r,\omega)=e^{-Zr}\mu(r,\omega)\) with
\(\mu(\,\cdot\,,\omega)\in C^{2,\alpha}([0,\infty))\).
It follows that
\(\rho(\,\cdot\,,\omega)\in C^{2,\alpha}([0,\infty))\)
since \(r\mapsto e^{-Zr}\)
belongs to
\(C^\infty([0,\infty))\). In particular,
\(\rho'(0,\omega)=\lim_{r\downarrow0}\rho'(r,\omega)\) and
\(\rho''(0,\omega)=\lim_{r\downarrow0}\rho''(r,\omega)\).
(All the above for all \(\alpha\in(0,1)\)).

Next, by the above, 
\begin{align*}
  \lim_{r\downarrow0}\rho'&(r,\omega)
  =\lim_{r\downarrow0}\big[\omega\cdot\nabla\rho(r,\omega)\big]
  \\&=\lim_{r\downarrow0}\big[-Z\rho(r,\omega)+e^{-Zr}
  \omega\cdot\nabla\mu(x)\big]  
  =-Z\rho(0)+\omega\cdot\nabla\mu(0),
\end{align*}
which is \eqref{eq:fixedW3}.

Finally, the proof of \eqref{eq:fixedW4}. Due to \(\rho=e^{-Zr}\mu\)
and \eqref{eq:1}--\eqref{nuatom}
we have
\begin{align*}
  \rho''(r,\omega)&=
  Z^2\rho(r,\omega)-2Ze^{-Zr}\omega\cdot\nabla\mu(x)
  \\&\quad+e^{-Zr}\big[\omega\cdot\nabla(\omega\cdot\nabla \nu(x)) +
  2C\cdot\omega\big].
\end{align*}
A simple computation shows that 
\(\omega\cdot\nabla(\omega\cdot\nabla
\nu(x)) =\omega\cdot \big((D^2\nu)(x)\omega\big)\), and, since \(\mu\in
C^{1,1}(\mathbb R^3)\) and 
\(\nu\in C^{2,\alpha}(\mathbb R^3)\), \(\alpha\in(0,1)\), we get
\eqref{eq:fixedW4}.  
\end{pf*}
\begin{pf*}{Proof of Theorem~\ref{thm:ThmC-2-alpha}}
We will show that the symmetry assumption
\eqref{symmetry} for \(\psi\) implies that
\(c_1=0\in\mathbb R^3\) in \eqref{eq:muNew1}. Then
\begin{align}
  \label{eq:muGood}
  \Delta\mu_1=g\  ,\  g\in C^\alpha(\mathbb R^3)\, , \, \alpha\in(0,1),
\end{align}
and so standard elliptic regularity implies that
\(\mu_1\in C^{2,\alpha}(\mathbb R^3)\) for all \(\alpha\in(0,1)\). 
This will prove \eqref{main3}.

Recall that
(see \eqref{eq:firstOrder}, \eqref{eq:claim}, \eqref{eq:muNew2},
\eqref{eq:hFinal}
\eqref{omegaInt1Bis},
\eqref{omegaInt2Bis}, and \eqref{eq:c-5-1}) 
\begin{align}
  \label{eq:cS}
  c_1&=c_1^{(1)}+2c_1^{(2)}=c_1^{(1)}+2c_{1,1}-4c_{3,1}+4c_{5,1},\\
    \label{eq:cSBis}
  c_1^{(1)}&=2Z\nabla\mu_1(0)\ , \ 
  c_{1,1}=\frac{Z}{4}{\rm Int}_1(0),\\
  \label{eq:cSBisBis}
  c_{3,1}&=-\frac{Z}{8}{\rm Int}_3(0)\ , \ 
  c_{5,1}=-\frac{Z}{4}{\rm Int}_5(0).
\end{align}

We first consider $c_1^{(1)}$. 
The assumption
\eqref{symmetry} clearly implies that $\rho_1$ is an even 
function on ${\mathbb R}^3$. It follows from \(\mu_1=e^{Z|x|}\rho_1\) 
that $\mu_1 \in C^{1,1}({\mathbb R}^3)$ is even and therefore 
$\nabla \mu_1 \in C^{0,1}({\mathbb R}^3; {\mathbb R}^3)$ is odd. In
particular, $\nabla \mu_1(0) =0$, and so \(c_1^{(1)}=0\).

It was shown in the proof of Theorem~\ref{thm:ThmC-1-1} that \({\rm
  Int}_j\in C^{\alpha}(\mathbb R^3;\mathbb R^3), j=1,3,5\),
\(\alpha\in(0,1)\) (see \eqref{int1}, \eqref{int3}, and
\eqref{int5}). Furthermore, the symmetry condition 
  \eqref{symmetry} clearly implies that all three functions are odd
  (for \({\rm Int}_5\), use \eqref{eq:psi1}). It follows that \({\rm
  Int}_j(0)=0, j=1,3,5\), and therefore (see \eqref{eq:cSBis} and
\eqref{eq:cSBisBis}) 
  \(c_{j,1}=0, j=1,3,5\). Therefore \(c_1^{(2)}=0\) and hence
  \(c_1=0\) in \eqref{eq:cS}. This, 
  and \eqref{eq:muNew1}, implies \eqref{eq:muGood}, which, as
  mentioned above, proves \eqref{main3}.
 
  Note that \(c_1^{(2)}=0\) implies that \(h\in C^{\alpha}(\R^3)\) for
  all \(\alpha\in(0,1)\) (see Lemma~\ref{lem:formH}).

 The above clearly implies that \(\nabla\mu(0)=0\) and
 \(C=\sum_{j=1}^Nc_j=0\), and so \eqref{eq:fixedW3} and
\eqref{eq:fixedW4} imply \eqref{cusp0}.

It remains to prove \eqref{eq:Marias}. With \(\Delta=\partial^2/\partial
r^2+(2/r)\partial/\partial r-\mathcal{L}^2/r^2\), \eqref{eq2:rho} becomes
(after multiplication by \(-r\))
\begin{align*} 
  r\rho''(r,\omega)+2\rho'(r,\omega)+2Z\rho(r,\omega)-2rh(r,\omega)=
  \frac{(\mathcal{L}^2\rho)(r,\omega)}{r}.
\end{align*}
This implies, using the fact that \(h\in L^{\infty}(\R^3)\)
and \eqref{cusp0}, that
\begin{align*}
  \lim_{r\downarrow0}  \frac{(\mathcal{L}^2\rho)(r,\omega)}{r}
  =2\big(\rho'(0,\omega)+Z\rho(0,\omega)\big)=0.
\end{align*}
Let \(\mathcal{R}(r,\omega):=
  \frac{(\mathcal{L}^2\rho)(r,\omega)}{r^2}\), then \eqref{eq2:rho}
  reads
  \begin{align}
    \label{eq:inPolar}
    \rho''(r,\omega)+\frac{2}{r}\big(\rho'(r,\omega)+Z\rho(r,\omega)\big)-2h(r,\omega) 
    =\mathcal{R}(r,\omega)\ , \ r>0.
  \end{align}
  Note that, by l'H\^{o}pital's rule and \eqref{cusp0},
  \begin{align*}
    \lim_{r\downarrow0}\frac{2}{r}\big(\rho'(r,\omega)+Z\rho(r,\omega)\big)
    =2\big(\rho''(0,\omega)+Z\rho'(0,\omega)\big),
  \end{align*}
and so \eqref{eq:inPolar} implies that
\(\mathcal{R}(0,\omega):=\lim_{r\downarrow0}\mathcal{R}(r,\omega)\)
exists, and 
\begin{align*}
  \mathcal{R}(0,\omega)=3\rho''(0,\omega)+2Z\rho'(0,\omega)-2h(0,\omega).
\end{align*}
The existence of \(h(0,\omega):=\lim_{r\downarrow0}h(r,\omega)\)
follows from Lemma~\ref{lem:formH}. 
Therefore, using \eqref{cusp0}, we obtain \eqref{eq:Marias}.
\end{pf*}

\appendix
\section{A useful lemma}\label{app:lem}
The following lemma is Lemma~2.9 in \cite{CMP2}; we include it, without
proof, for the convenience of the reader. (The proof is simple, and
can be found in \cite{CMP2}). 
\begin{lemma}\label{usefullemma}
  \label{lem:XdotG}
  Let \(G:U\to\mathbb R^{n}\) 
for \(U\subset\mathbb R^{n+m}\) a neighbourhood of
  a point \((0,y_{0})\in\mathbb R^{n}\times\mathbb R^{m}\). Assume
  \(G(0,y)=0\) for 
  all \(y\) such that \((0,y)\in U\). Let
  \begin{align*}
    f(x,y)=\left\{\begin{array}{cc}
             \frac{x}{|x|}\cdot G(x,y)& x\neq 0, \\
             0& x=0. \\
  \end{array}\right. 
  \end{align*}
  Then, for \(\alpha\in(0,1]\), 
  \begin{align}
    \label{eq:lem_G=0}
     G\in C^{0,\alpha}(U;\mathbb R^{n})\Rightarrow f\in C^{0,\alpha}(U).
  \end{align}
Furthermore, $\| f \|_{C^{\alpha}(U)} \leq 2\| G \|_{C^{\alpha}(U)}$.
\end{lemma}

\begin{acknowledgement}
Parts of this work have been carried out at various
institutions, whose hospitality is gratefully acknowledged:
Mathematisches Forschungs\-institut Ober\-wolfach 
(SF, T\O S), The Erwin Schr\"{o}\-dinger Institute (SF, T\O
S), Universit\'{e} Paris-Sud (T\O S), and the IH\'ES (T\O S).
Financial support from the 
European Science Foundation Programme {\it Spectral Theory and Partial 
  Differential Equations} (SPECT), and EU IHP network 
{\it Postdoctoral Training Program in Mathematical Analysis of
Large Quantum Systems},
contract no.\
HPRN-CT-2002-00277, is
gratefully acknowledged.
T\O S was partially supported
by the embedding grant from The Danish National
Research Foundation: Network in Mathematical Physics and Stochastics, and
by the European Commission through its 6th Framework Programme
{\it Structuring the European Research Area} and the contract Nr.
RITA-CT-2004-505493 for the provision of Transnational Access
implemented as Specific Support Action.
\end{acknowledgement}

\bibliographystyle{amsplain}

\providecommand{\bysame}{\leavevmode\hbox to3em{\hrulefill}\thinspace}
\providecommand{\MR}{\relax\ifhmode\unskip\space\fi MR }
\providecommand{\MRhref}[2]{%
  \href{http://www.ams.org/mathscinet-getitem?mr=#1}{#2}
}
\providecommand{\href}[2]{#2}

\end{document}